\RequirePackage{fix-cm}
\documentclass[twocolumn,epjc3]{svjour3}
\smartqed  
\RequirePackage{graphicx}
\RequirePackage[caption=false]{subfig}
\RequirePackage{xcolor}
\usepackage{bm}
\usepackage{xspace}
\usepackage{multirow}
\usepackage{xcolor}
\usepackage[british]{babel}
\usepackage{hyphenat}
\usepackage{amsmath,amssymb}

\usepackage{graphicx} 
\usepackage{caption}
\usepackage{subfig}
\usepackage{float} 
\graphicspath{{figures/}} 
\usepackage{booktabs}
\usepackage{hyperref}

\newcommand{\beq}{\begin{equation}}
\newcommand{\eeq}{\end{equation}}

\journalname{Eur. Phys. J. C}
\begin{document}

\title{Muon identification in a compact single-layered water Cherenkov detector and gamma/hadron discrimination using Machine Learning techniques 
}


\author{R. Concei\c{c}\~ao\thanksref{addr1,addr2}
        \and
        B.S. Gonz\'alez \thanksref{addr1,addr2,addr3}
        \and
        A. Guill\'en\thanksref{addr3}
        \and
        M. Pimenta\thanksref{addr1,addr2} 
        \and \\
        B. Tom\'e\thanksref{addr1,addr2}
}

\thankstext{e1}{e-mail: borjasg@lip.pt}


\institute{
LIP, Av. Prof. Gama Pinto, 2, P-1649-003 Lisbon, Portugal \label{addr1}
\and
 Instituto Superior T\'{e}cnico (IST), Universidade de Lisboa, Av. Rovisco Pais 1, 1049-001, Lisbon, Portugal \label{addr2}
\and
 Computer Architecture and Technology Department, University of Granada, Granada, Spain \label{addr3}
}

\date{Received: date / Accepted: date}

\maketitle

\begin{abstract}

The muon tagging is an essential tool to distinguish between gamma and hadron-induced showers in wide field-of-view gamma-ray observatories. In this work, it is shown that an efficient muon tagging (and counting) can be achieved using a water Cherenkov detector with a reduced water volume and 4 PMTs, provided that the PMT signal spatial and time patterns are interpreted by an analysis based on Machine Learning (ML). 
The developed analysis has been tested for different shower and array configurations. The output of the ML analysis, the probability of having a muon in the WCD station, has been used to notably discriminate between gamma and hadron induced showers with $S/ \sqrt{B} \sim 4$ for shower with energies $E_0 \sim1\,$TeV. Finally, for proton-induced showers, an estimator of the number of muons was built by means of the sum of the probabilities of having a muon in the stations. Resolutions about $20\%$ and a negligible bias are obtained for vertical showers with $N_{\mu} > 10$.

\keywords{High Energy gamma rays\and Wide field-of-view observatories \and Water Cherenkov Detectors \and Muon identification  \and Gamma/hadron discrimination \and Machine Learning}
\end{abstract}

\section{Introduction}

The study of gamma-rays is crucial to investigate our surrounding Universe. Their neutral nature allows them to cover long distances along the Universe without being deflected by magnetic fields. Thus, the detection of gamma-rays can be used to track emitting astrophysical sources. In particular, gamma-rays ranging from $\sim 100$ GeV and few hundred TeV, known as very high-energy (VHE) gamma-rays, are very interesting to investigate some of the most extreme non-thermal events taking place in the Universe. Active Galactic Nuclei (AGNs) --supermassive blackholes in the center of galaxies, powered by infalling matter-- and Gamma Ray Bursts (GRBs) --intense and fast shots of gamma radiation-- are examples of interesting target sources, both multi-messenger astrophysical counterparts of VHE neutrinos and gravitational wave events \cite{IceCube18,Abbott17a,Abbott17b}.
Furthermore, not only is gamma-rays detection important for identifying astrophysical emitting sources, but it could also be essential to prove the existence of new physics at fundamental scales beyond the standard model of particle physics \cite{Pimenta2018Astroparticle} and to provide answers to some fundamental open questions in physics, for instance, the nature of dark matter \cite{viana2019searching}.

The direct detection of gamma-rays is only possible using satellite-borne instruments \cite{atwood2009large}. Nevertheless, above a few hundreds of GeV, the gamma-ray flux becomes too small and only ground-based experiments can indirectly detect this kind of radiation. Indirect detection techniques take advantage of the secondary shower of particles, known as Extensive Air Showers (EAS), produced by the interaction of the gamma-ray with the Earth's atmosphere to infer their direction and energy \cite{degrange2015introduction}. However, even though indirect methods are very effective at the VHE energy region, a significant drawback is that one has to deal with an enormous hadronic background produced by the cosmic-rays continuously reaching the Earth \cite{albert2019science}. At the sub-TeV region, the rejection of cosmic-rays can be done by exploring the patterns of the secondary particles reaching the ground \cite{bartoli2013tev,assunccao2019automatic}.
At the TeV energy range, muons begin to appear in significant quantities in hadronic showers. Therefore, the detection of muons provides an important tool to efficiently separate between gamma and hadronic extensive air showers  \cite{westerhoff2014hawc,zuniga2017detection}.

This work aims to demonstrate that a dedicated design of the water Cherenkov detector (WCD) combined with state-of-the-art machine learning techniques can be used to identify muons and with it provide an effective gamma/hadron discrimination. Such a detector can have a reduced water height while maintaining a high physics performance, making it suitable for large experiments placed at extreme altitudes (around 5 km a.s.l.).

Hence, for the present study, we propose the use of an EAS wide field-of-view array composed of water Cherenkov detectors (Section \ref{sec:WCD_conf}). The signals time traces recorded by the WCDs will be analysed to distinguish between muons and electromagnetic shower secondary particles (electrons, positrons and photons) by means of a \textit{Convolutional Neural Network} (Section \ref{sec:muon_disc}). Afterwards, the information about muons found by the algorithm was used to carry out a gamma/hadron discrimination (Section \ref{sec:gamma_hadron}) and estimate the number of muons in hadronically induced showers (Section \ref{sec:muon_counting}). Finally, a discussion on the performance for inclined shower events and a sparser detector array is presented in Section \ref{sec:discussion}, followed by the conclusions.

\section{WCD configuration} \label{sec:WCD_conf}

Several approaches to identify muons in ground-based wide field-of-view gamma-ray experiments have been tried or suggested over time. One of the most successful approaches comes from the HAWC experiment \cite{HAWC_tank}. The HAWC tanks contain a large volume of water and black walls. This can be used to attenuate the electromagnetic particles whilst muons traverse the whole detector giving a large signal and an unmistakable signature. Although highly successful on the TeV energy range, this detector has a limited performance in the sub-TeV energies, and its massive amount of water does not make it a good candidate to place it at a higher altitude (one of the most effective ways to lower the energy threshold).
 
Another option would be to use dedicated muon detectors, buried or shielded, as it is being done in LHAASO \cite{LHAASO_muon} or simply below the water tank, as proposed in the MARTA\footnote{although MARTA was primarily planned to be used in the study of Ultra-High Energy Cosmic Rays} project \cite{MARTA}. These hybrid experiments have many advantages as the shower components can be scrutinised by independent detectors. However, it also dramatically increases the cost of the project. Note that VHE gamma-ray experiments such as LHAASO or the future Southern Wide-field Gamma-ray Observatory (SWGO) \cite{SWGO,abreu2019southern}, need to cover huge areas up to $1\,{\rm km^2}$ and muons are relatively scarce for $\sim\,$TeV showers.

\begin{figure}[t]
 \centering
 \includegraphics[width=0.30\textwidth]{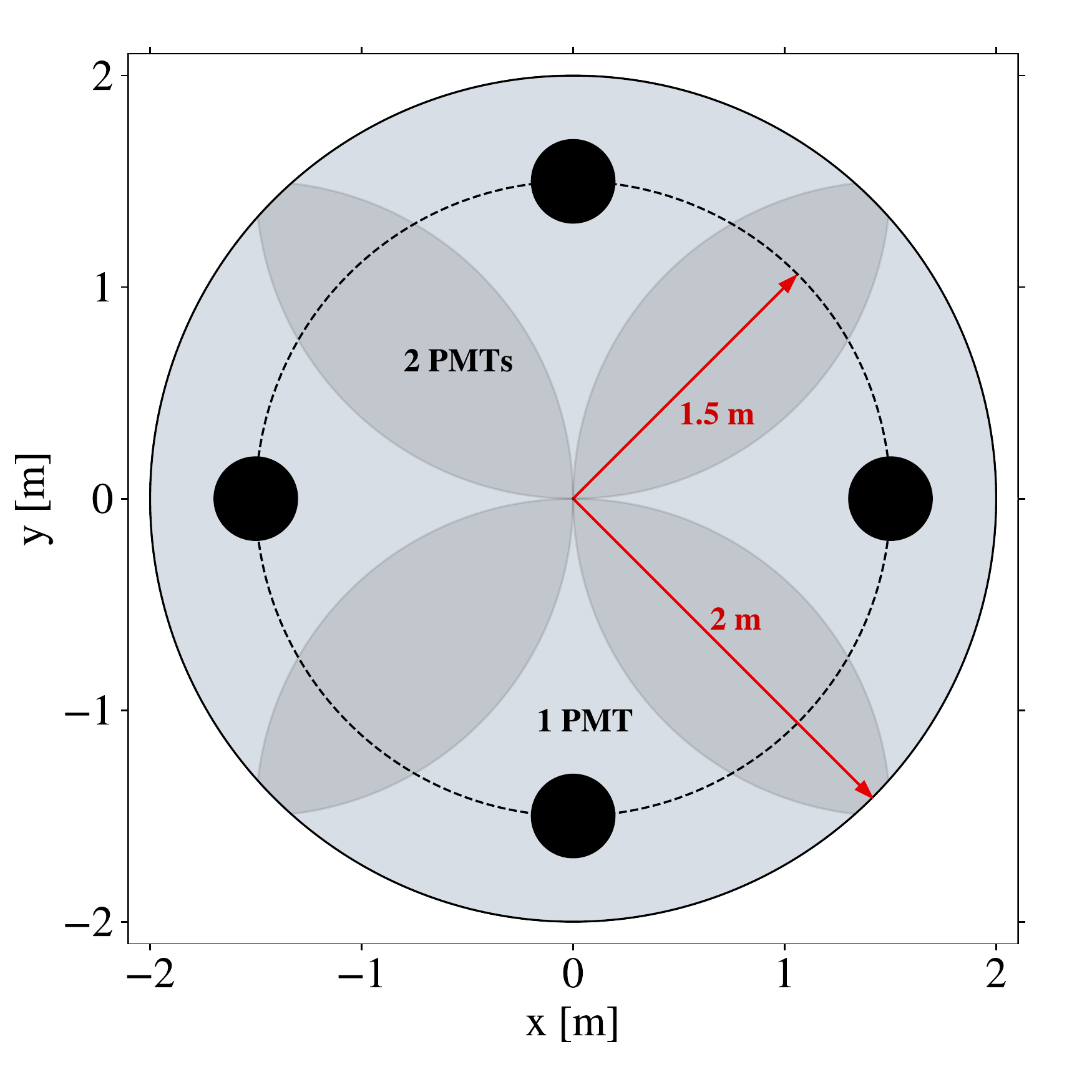}    
 \caption{Single-layered WCD design. A dark circular area with a radius of 1.5 m was drawn around each 8-inch PMT (represented as black circles). Taking into account the height of this WCD, the direct Cherenkov light of a vertical muon that crosses the WCD through one of these areas should be detected by its correspondent PMT (it could be detected by two PMTs in the case that a muon crosses the station through the intersection of two dark areas).}
 \label{fig:WCD_design}
\end{figure}

In this work, we aim to explore another via that uses a single-layered WCD with 4 PMTs (see Figure \ref{fig:WCD_design}). The rationale behind this concept is that muons will cross the whole detector, and as such, the direct Cherenkov light, produced along with its traversal, will reach only a portion of the WCD floor. Contrary, photons and electrons give origin to electromagnetic showers inside the station, creating a broader Cherenkov light pool. Moreover, muons travel mostly alone while electromagnetic particles arrive in bulks, further supporting this signal asymmetry view.

\begin{figure}[t]
 \centering
 \includegraphics[width=0.49\textwidth]{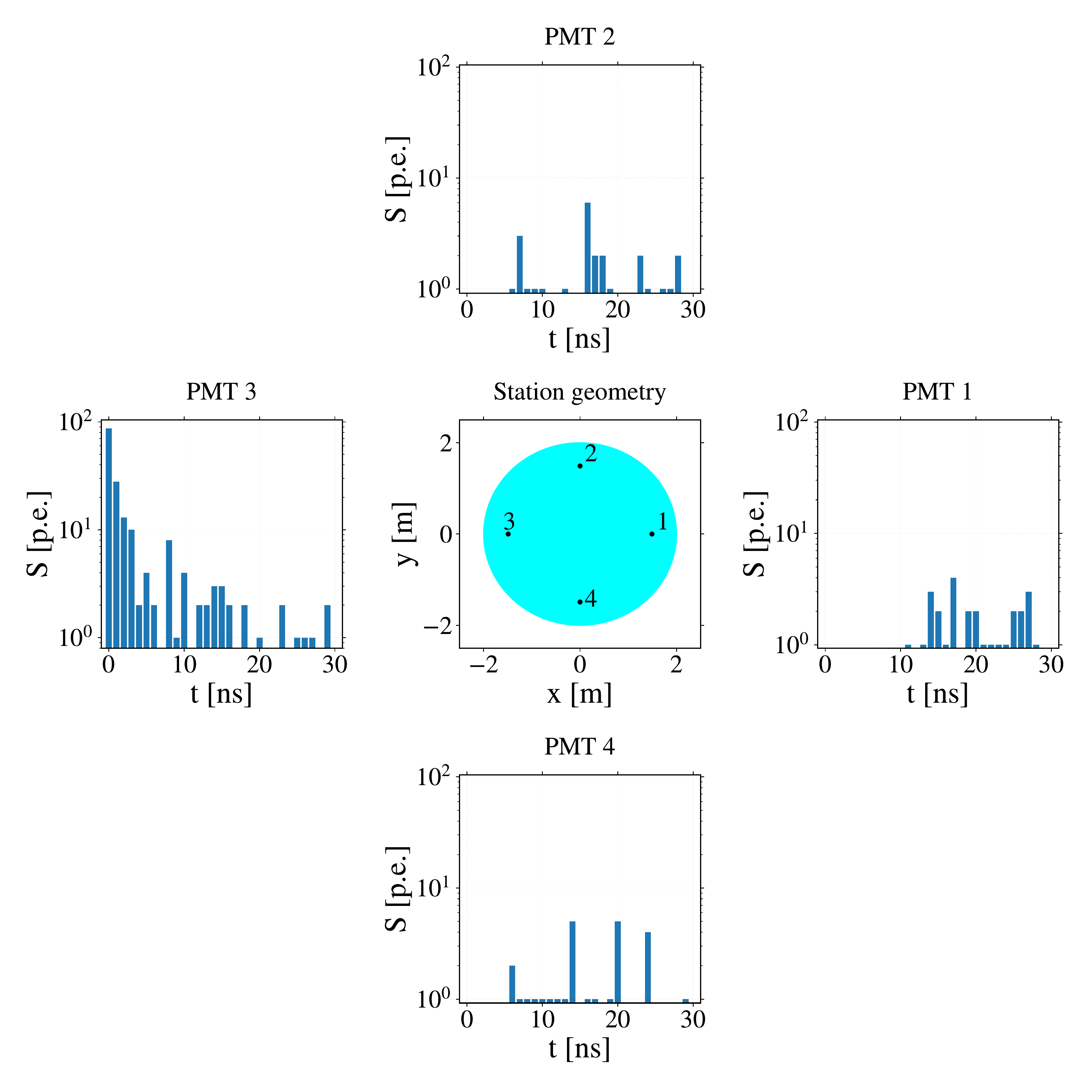}    
 \caption{Single-layered WCD crossed by a single relativistic muon. The WCD drawing is surrounded by the correspondent PMT signal time traces.}
 \label{fig:WCD_and_trace}
\end{figure}

The station was designed such that the signal asymmetry caused by a vertical muon is maximal while ensuring a complete signal coverage (signal uniformity). The area of the stations is a parameter that should be optimised, taking into account the array's physics performance and the station cost. On the one hand, the higher the segmentation, the higher the shower ground pattern detail (and consequent improvement of the physics analyses). On the other hand, the bigger the station, the lower the cost (fewer materials and photo-sensors per area). In this work, we have considered cylindrical tanks with a base diameter of 4 m as a reasonable compromise (for reference, the LHAASO-WCDA has an area of $25\,{\rm m^2}$).

For stations with a base diameter of 4 m, the water height and the distance of the PMTs to the tank centre, $r_{\rm PMT}$, will depend on the number of PMTs considered. For 3, 4 and 5 PMTs, we would get an optimal distance of the PMTs to the centre of the tank of 2 m, 1.5 m and 1.25 m, respectively and a corresponding water height of 2.3 m, 1.7 m and 1.25 m. The parameter $r_{\rm PMT}$ is chosen by placing the PMTs at a position such that together they can always collect the light produced by a vertical entering muon in any position of the tank while minimising the overlap between PMTs (see Figure~\ref{fig:WCD_design} of the paper).

The water height is obtained, noting that we now aim for a Cherenkov light pool of a radius $r_{\rm opt}$ and the Cherenkov angle of a relativistic muon in water is roughly $\sim 41^\circ$.
The higher the number of PMTs, the lower the needed water height. The exact number of PMTs and thus of the station dimensions will depend on the optimisation of several parameters, which is out of the scope of the paper. They will depend on the cost of PMTs and the cost (availability) of water at very high altitude.

In this work, we aim to show that the tagging of muons through the signal time trace asymmetry is a viable option. For that, we have considered four PMTs the corresponding stated above dimensions. Given the arguments before, the paper results should be valid for a slightly lower/higher number of PMTs, provided that the dimensions of the WCD scale accordingly.

The walls of this WCD should be white diffusive to maximise the signal collection. By construction, the WCD should have a good light collection uniformity independently of the muon's entry point. Both features are essential to effectively lower the energy threshold.

The shower geometry reconstruction is done through the arrival time of the shower front. To reach adequate angular resolutions, the shower plane should be measured with time resolutions of the order of a few nanoseconds~\cite{HAWC_tank,LATTES}. With the proposed design, such can easily be achieved by exploiting the first peak in the PMT signal time trace, associated with the direct Cherenkov light production (see for instance, the signal time trace of PMT 3 in Figure \ref{fig:WCD_and_trace}). The directly light peak appears typically within the first $10\,$ns to the signal starting time trace, $T_0$, and has a width of $2-3$\,ns. 

In summary, in this work, we shall exploit a water Cherenkov detector with the following characteristics:
\begin{itemize}
\item Tank diameter $4\,$ m;
\item Tank height $1.7\,$m;
\item PMT distance to the WCD centre of $1.5\,$m.
\item White diffusive walls
\item Four photomultipliers with 8 inches of diameter
\end{itemize}

As qualitatively argued before, the proposed WCD station can trigger and reconstruct the shower energy and geometry. However, to make it appropriate for the study of gamma-rays, we need to be able to use it to identify/count EAS muons. Such will be the focus of the following sections.

\section{Simulation and analysis strategy} \label{sec:simulation}
\subsection{Simulations}

The extensive air showers used in this work were simulated with CORSIKA (version 7.5600) \cite{CORSIKA} and the detector response using the Geant4 toolkit (with version 4.10.05.p01) \cite{agostinelli2003geant4,Geant4_2006,Geant4_2016}. 

To train and assess the machine learning algorithms' performance, it was used proton-induced shower simulations with energies $E_0 \in [4;6]\,$TeV. This energy range was selected to ensure a high number of stations with only muons and no electromagnetic contamination nor any other particles crossing the WCD. Moreover, it was shown to be desirable for the training not to have more than one muon in the WCD. The showers were generated following a $E_0^{-1}$ spectra, an azimuth angle uniformly distributed, and a zenith angle $\theta_{0} \in [5^\circ;15^\circ]$ (vertical events) and $\theta_{0} \in [25^\circ;35^\circ]$ (inclined events). 

The experiment observation level was set at $5\,200\,$m above the sea level\footnote{This altitude corresponds to the altitude of the ALMA site, Chile, one of the sites being currently considered for SWGO.}. In the energy range used for this study ($\sim  1$ TeV), the number of muons at lower altitudes
would be essentially the same, and only the electromagnetic shower component would be attenuated
(effectively increasing the energy threshold). Thus, the method to tag/count muons in the WCD proposed in this work is expected to work with similar performance regardless of the experiment's altitude (provided that it is above something like $4\,000\,$m a.s.l.).

Two array configurations of WCDs are compared to investigate the impact of the WCDs distribution at ground in the results for inclined showers. Their parameters were chosen for being one of the possibilities for SWGO. The array matrices cover an area of roughly $80\,000 \ m^2$ and have the following characteristics:

\begin{itemize}
\item a dense array which is composed of $5\,720$ WCDs nearly touching each other (fill factor of $\sim 80\%$);
\item a sparse array with $1\,906$ WCDs with a minimum separation of $\approx 12\,$m. This array was built by removing stations such that no WCD has a neighbour station closer than its own diameter.
\end{itemize}

\begin{figure}[t]
 \centering
 \includegraphics[width=0.4\textwidth]{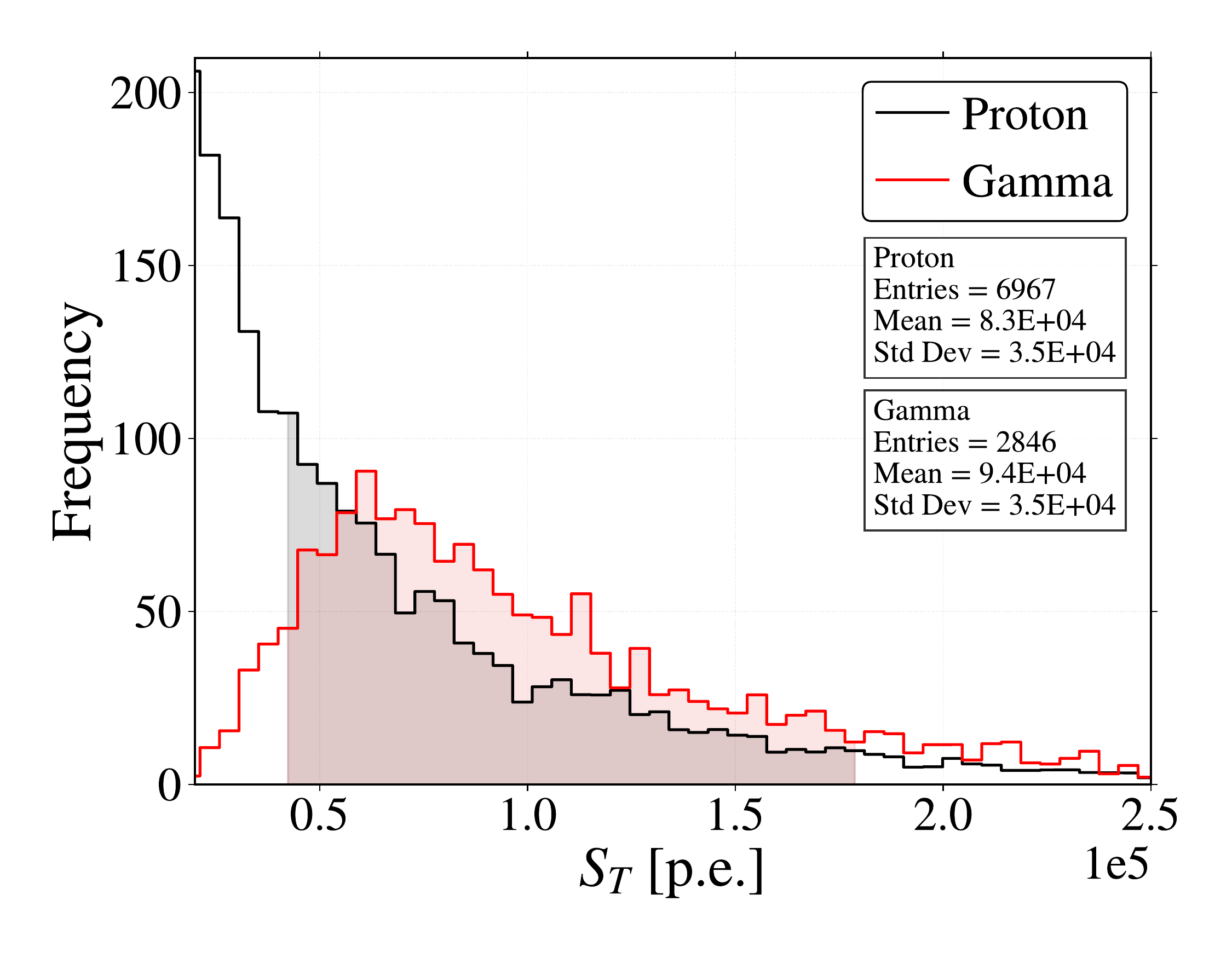}    
 \caption{Signal at the ground of gamma and proton induced EAS used in Section \ref{sec:gamma_hadron}. The area below the selected events was filled in red for gammas and grey for protons. The statistical data showed in this figure corresponds only to the selected events.}
 \label{fig:signal_ground}
\end{figure}

In Sections \ref{sec:gamma_hadron} and \ref{sec:muon_counting}, the identification of muons is used to perform the gamma/hadron discrimination and to estimate the number of muons in hadronic showers. A particular set of simulations was derived in order to enable a sensible comparison between gamma and hadron events. Gamma-induced showers were generated with energies between $E_0 \in [1;1.6]\,$TeV, while proton-induced showers were simulated for energies between $\sim 600\,$GeV and $6\,$TeV. The zenith angles for both are comprised between 5$^{\circ}$ -- 15$^{\circ}$. Afterwards, a cut on the total measured WCD signal at the ground is applied to emulate a typical energy reconstruction (see Figure \ref{fig:signal_ground}): 

$$S_T \in \left [ \left \langle S_{T, \gamma} \right \rangle - \sigma_{S_{T, \gamma}} ; \left \langle S_{T, \gamma} \right \rangle + \sigma_{S_{T, \gamma}}  \right ] = \left [ 0.4 ; 1.8  \right ] \ \cdot \ 10^{5} \ \mathrm{p.e.}$$

where $S_T$ is the total signal at ground using all the WCDs, $\left \langle S_{T, \gamma} \right \rangle$ and $\sigma_{S_{T, \gamma}}$ are the average and standard deviation respectively of the distribution for $S_T$ in gamma-ray events. Since the events were generated with a $E_0^{-1}$ spectra, the distributions were weighted before applying the cut on the signal and during the entire experimentation in Section \ref{sec:gamma_hadron}. The weights $E_{0, \gamma}^{-1}$ and $E_{0, p}^{-2}$ were applied to ensure a realistic power law spectrum of energies (expected to be $\sim E_{0, \gamma}^{-2}$ and $\sim E_{0, p}^{-3}$). 

To sum up, in Tables \ref{tab:training_description} and \ref{tab:datasets_description} we describe the data sets used to train and assess the artificial neural networks. It should be noted that some considerations were taken into account before building these data sets. To avoid events where the muon partially crossed the station, only stations with more than 300 photoelectrons (p.e.) were used during the experimentation (train and test of the machine learning algorithms). Only \textit{Single Muons}\footnote{WCD with only one muon crossing it and nothing else} (as well as stations without muons) were considered during the training process in order to provide representative muon samples to the algorithm. All kind of stations with muons are considered for the test sets.  A second training data set was built to train again the neural network with inclined events (using the same configuration of the algorithm that was optimised for vertical events).

\begin{table}[H]
\centering
\scriptsize
\begin{tabular}{@{}cccc@{}}
\cmidrule(l){2-4}
\multirow{2}{*}{} & \multicolumn{2}{c}{$\theta_ 0 \in \left[ 5^{\circ}; 15^{\circ} \right]$} & $\theta_ 0 \in \left[ 25^{\circ}; 35^{\circ} \right]$ \\ \cmidrule(l){2-4} 
                  & Training                            & Validation                         & Training                                              \\ \midrule
EAS               & $2\,000 \ (80 \%)$                    & $2\,000 \ (20 \%)$                   & 655                                                   \\
S.M. stations      & $13\,795$                           & $3\,449$                           & $5\,472$                                              \\
E.M. stations       & $272\,006$                          & $68\,001$                          & $85\,130$                                             \\
Instances         & $285\,801$                          & $71\,450$                          & $90\,602$                                             \\
Muonic prop.      & 4.83 $\%$                           & 4.83 $\%$                          & 6.04 $\%$                                             \\ \bottomrule
\end{tabular}
\caption{Number of stations with \textit{Single Muons} (S.M. stations) and without muons (E.M. stations) of the training data sets. Data sets created using vertical ($\theta_ 0 \in \left[ 5^{\circ}; 15^{\circ} \right]$) and inclined ($\theta_ 0 \in \left[ 25^{\circ}; 35^{\circ} \right]$) showers with primary energy $E_0 \in \left[ 4; 6 \right]$ TeV and the dense array. A validation data set was created using a $20 \%$ of the stations from the original $2\,000$ vertical EAS.}
\label{tab:training_description}
\end{table}

\begin{table}[H]
\centering
\scriptsize
\resizebox{8.4cm}{!} {
\begin{tabular}{@{}ccccccc@{}}
\toprule
Primary & Array  & $E_0$ (TeV)             & $\theta_{0}$ ($^{\circ}$) & EAS & \begin{tabular}[c]{@{}c@{}}Stations\\ with All Muons\end{tabular} & \begin{tabular}[c]{@{}c@{}}Stations \\ without muons\end{tabular} \\ \midrule
Proton  & Dense  & $\left[ 4; 6 \right]$   & $\left[ 5; 15 \right]$                                              & $1\,693$                                                         & $57\,283$                                                        & $289\,411$                                                   \\

Proton  & Dense  & $\left[ 4; 6 \right]$   & $\left[ 25; 35 \right]$                                             & 281                                                          & $7\,601$                                                        & $34\,738$  \\
Proton  & Sparse & $\left[ 4; 6 \right]$   & $\left[ 25; 35 \right]$                                             & 936                                                          & $8\,452$                                                         & $40\,933$                                                    \\

Proton  & Dense & $\left[ 0.6; 6 \right]$ & $\left[ 5; 15 \right]$                                              & $6\,967$                                                         & $107\,579$                                                       & $379\,421$                                                   \\
Gamma   & Dense & $\left[ 1; 1.6 \right]$   & 10                                                                  & $2\,846$                                                         & 263                                                          & $174\,103$                                                   \\ \bottomrule
\end{tabular}}
\caption{Description of the test data sets.}
\label{tab:datasets_description}
\end{table}

\subsection{Analysis strategy}
The objective function defined for this problem is the probability, $P_{\mu}^{(i)} \in \left[ 0, 1 \right]$, that a muon has passed through the WCD $(i)$ whose signal is being analysed. In such a way, the method is intended to be able to tag contaminated events where not only muons have crossed through the station.

To carry out this identification of stations with muons we propose the use of variables based solely on the WCD signal \cite{BorjaOtherPaper}. With the following variables we aim to explore both temporal (patterns in the signal time traces) and spatial (asymmetry in the PMTs' integrals) features:

\begin{itemize}
    \item Normalised signal time trace of each PMT. 
    \item Integral of each PMTs signal time trace.
    \item Cherenkov light measured in the WCD (sum of the four PMTs' signal trace integrals).
    \item Normalised integral of each PMTs signal time trace.
\end{itemize}

The normalisation was done using the total Cherenkov light measured by the four PMTs during the time window considered for the variable that is going be normalised. Note that the normalised signal traces contain the first 30 nanoseconds to explore features from both direct the Cherenkov light and the reflections, while the rest of variables considered only the direct Cherenkov light (first 10 nanoseconds).

\subsection{Convolutional Neural Networks: design, training and optimisation}
Convolutional Neural Networks (CNNs) have become a standard in many machine learning applications. CNNs are feed-forward Artificial Neural Networks which combine two main blocks of hidden layers: \textit{convolutional layers} and \textit{fully-connected layers}. By using these layers, the CNN is able to fuse the processes of extracting features from data and perform a classification or regression.
In this work we use a 1-dimensional Convolutional Neuronal Network to extract complex features from the signals traces of the PMTs and combine it with the spatial features (integrals). With it, we define a regression problem to provide the probability $P_{\mu}^{(i)}$.  

The configuration of the algorithm was optimised evaluating possible values with the validation data set for vertical showers. One input channel of data was established for each PMT signal trace. Three convolutional layers were set to study the signal time traces. ReLU activation function is used and the number of filters in these layers was 20, 15 and 10. The size of the filters was established by taking into consideration the mean signal traces of the stations muons, which reveals that the first pulse of direct Cherenkov light usually appears within the first 2 nanoseconds. Therefore, small filters of size 2 are introduced, including a stride equal to two in the first filter to emphasise the first maximum of signal. Afterwards, three dense layers are introduced to perform the regression using the previous signal features and the spatial variables. These layers are composed of 30, 15 and 10 neurons respectively. A final output layer with a single neuron was used to compute the final probability.  Sigmoid activation function was used for the three dense layers and final neuron.  Adam learning algorithm was used during the training stage, which is a stochastic gradient-based method broadly applied in many deep learning applications. The adjustment of its parameters was done following those recommended in \cite{kingma2014adam}, that is: $(\beta_1, \beta_2$) = (0.9, 0.999) and $\varepsilon$ = $10^{-7}$. Finally, the model was trained during 200 epochs with a batch size of 512 and a learning rate of $10^{-3}$. Python 3.7 and Keras were used as the framework of the entire study.

Note that the stations with muons constitute roughly a $5 \ \%$ of the total (see Table \ref{tab:training_description}). Therefore, the class ratio must be balanced before training \cite{Gonzalez_2020}. A random oversampling technique was applied to the training set, creating new samples of stations with muons by randomly repeating those available in the data set. Finally, the class ratio was adjusted to $N_\mu / N_{e.m.} = 0.5$. The class ratio was not completely balanced in order to avoid false positives.

\section{Discrimination of muons in the WCD} \label{sec:muon_disc}

To assess the performance of the method, the probability $P_{\mu}^{(i)}$ will be measured in stations with and without muons. Since the algorithm was trained using only Single Muons, the stations with muons will be distributed in different subclasses to study the impact of the electromagnetic contamination in the probability. 

The simulation only provided the amount of signal produced in the WCD ($S_{T}$), but not the two components ($S_{e.m.}$ and $S_{\mu}$) separately as it does with the energy. Therefore, a relation between the signal produced in events with Single Muons and its energy was calibrated. This allowed us to estimate the signal produced by the muon by means of its energy. Afterwards, the proportion of electromagnetic induced signal $s_{e.m.}\in \left[ 0;1 \right]$ was computed using the equation (\ref{eq:contamination}).

\begin{equation} 
    s_{e.m.} = \frac{S_{T}-S_{\mu}}{S_{T}} = \frac{S_{e.m.}}{S_{T}} 
    \label{eq:contamination}
\end{equation}

\begin{figure}[t]
 \centering
 \includegraphics[width=0.4\textwidth]{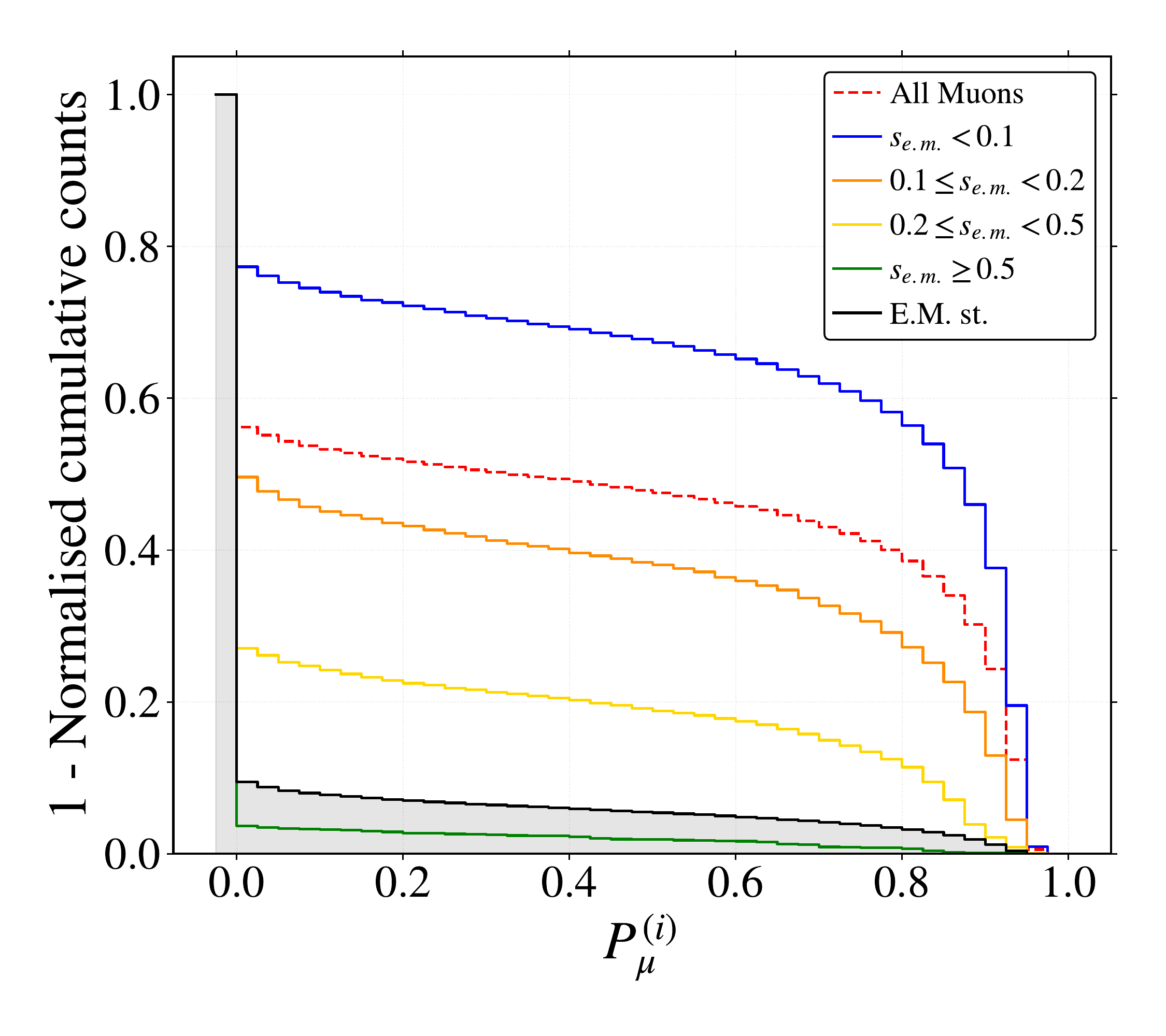}    
 \caption{Impact of electromagnetic contamination using the normalised inverse cumulative function for the probability $P_{\mu}^{(i)}$. Stations from proton-induced events with $E_0 \in \left[4,6 \right]$ TeV and $\theta_0 \in \left[ 5^{\circ} ; 15^{\circ} \right]$.}
 \label{fig:hist_contamination}
\end{figure}

In Figure \ref{fig:hist_contamination} it is shown the normalised inverse cumulative function for the probability $P_{\mu}^{(i)}$ measured in stations from proton induced events with $E_0 \in \left[4,6 \right]$ TeV and $\theta_0 \in \left[ 5^{\circ} ; 15^{\circ} \right]$. Most events with muons and low electromagnetic contamination were correctly identified, where roughly a $70 \%$ have a probability $P_{\mu}^{(i)} \geq 0.5$. For higher electromagnetic contamination we get lower $P_{\mu}^{(i)}$ values. Approximately a $40 \%$ and a $ 20 \%$ of the events with $s_{e.m.} \in \left[0.1;0.2 \right]$ and $s_{e.m.} \in \left[0.2;0.5 \right]$ respectively were identified with $P_{\mu}^{(i)} \geq 0.5$. Almost no probability is attributed to those events where most of the signal was not originated by the muon. This low probability may be justified given that these events involve a high electromagnetic signal, which added to the one caused by the muon, can cause a lower asymmetry among the PMTs signals. Taking into account all possible stations with muons, around a $50 \%$ of the stations had $P_{\mu}^{(i)} \geq 0.5$. This result proves that the algorithm was able to learn the main characteristics of the muonic signal from the \textit{Single Muons} events, so that the neural network is able to identify most single muons events and some contaminated events that share some of their features with single muons. Finally, less than a $10\%$ of the stations without muons had a probability $P_{\mu}^{(i)} > 0$.

\section{Gamma/hadron discrimination strategies} \label{sec:gamma_hadron}

In this section, we propose a gamma/hadron discrimination strategy which relies on the use of solely the information extracted in the previous sections at the WCD station level, $P_\mu^{(i)}$. This information should be combined at the shower event. A simple and intuitive observable was created for this purpose in equation (\ref{eq:prob}).

\begin{equation} \label{eq:prob}
    P_{\mu} = \sum_{i=1}^{N_S} P_{\mu}^{(i)}
\end{equation}

As seen before, $P_\mu^{(i)}$ is the probability, derived by the ML method, that the WCD station $i^{\rm th}$ was crossed by a muon, and $N_S$ is the total number of WCD stations with more than 300 p.e. in the shower event.

\begin{figure}[t]
 \centering
 \includegraphics[width=0.4\textwidth]{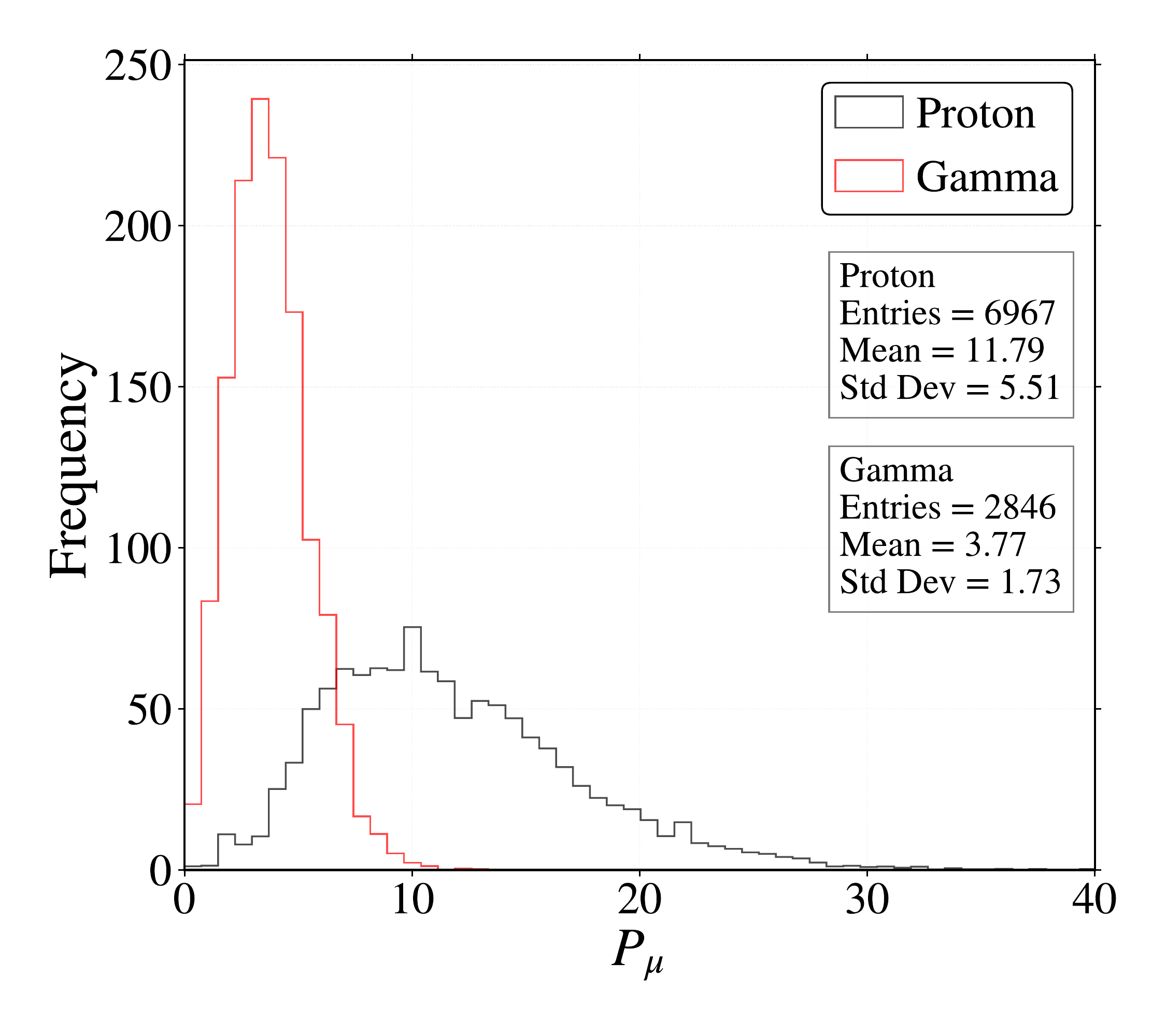}    
 \caption{Distribution of the probability $P_\mu$ after evaluating all the stations of the array for gamma and proton induced showers with energies $E_0 \in [1;1.6]\,$TeV and $E_0 \in [0.6;6]\,$TeV respectively.}
 \label{fig:prob_stations}
\end{figure}

The observable established in equation (\ref{eq:prob}) is tested for proton and gamma induced showers in Figure \ref{fig:prob_stations}. As expected, larger values are found when evaluating those events which have muons, i.e., proton initiated air showers. Moreover, it can be seen that by adequately choosing a cut on $P_\mu$, the gamma-induced showers can be efficiently separated from the ones generated by protons. It should be noted that there could be proton events of energy $E_0 \sim 0.6$ TeV with few or without muons, leading to low probabilities $P_\mu$.

A proxy to a gamma-ray experiment flux sensitivity can be obtained evaluating $S/\sqrt{B}$, where \textit{S} and \textit{B} are the selection efficiency for gamma-rays and the background (protons induced events), respectively. 

\begin{figure}[t]
 \centering
 \includegraphics[width=0.4\textwidth]{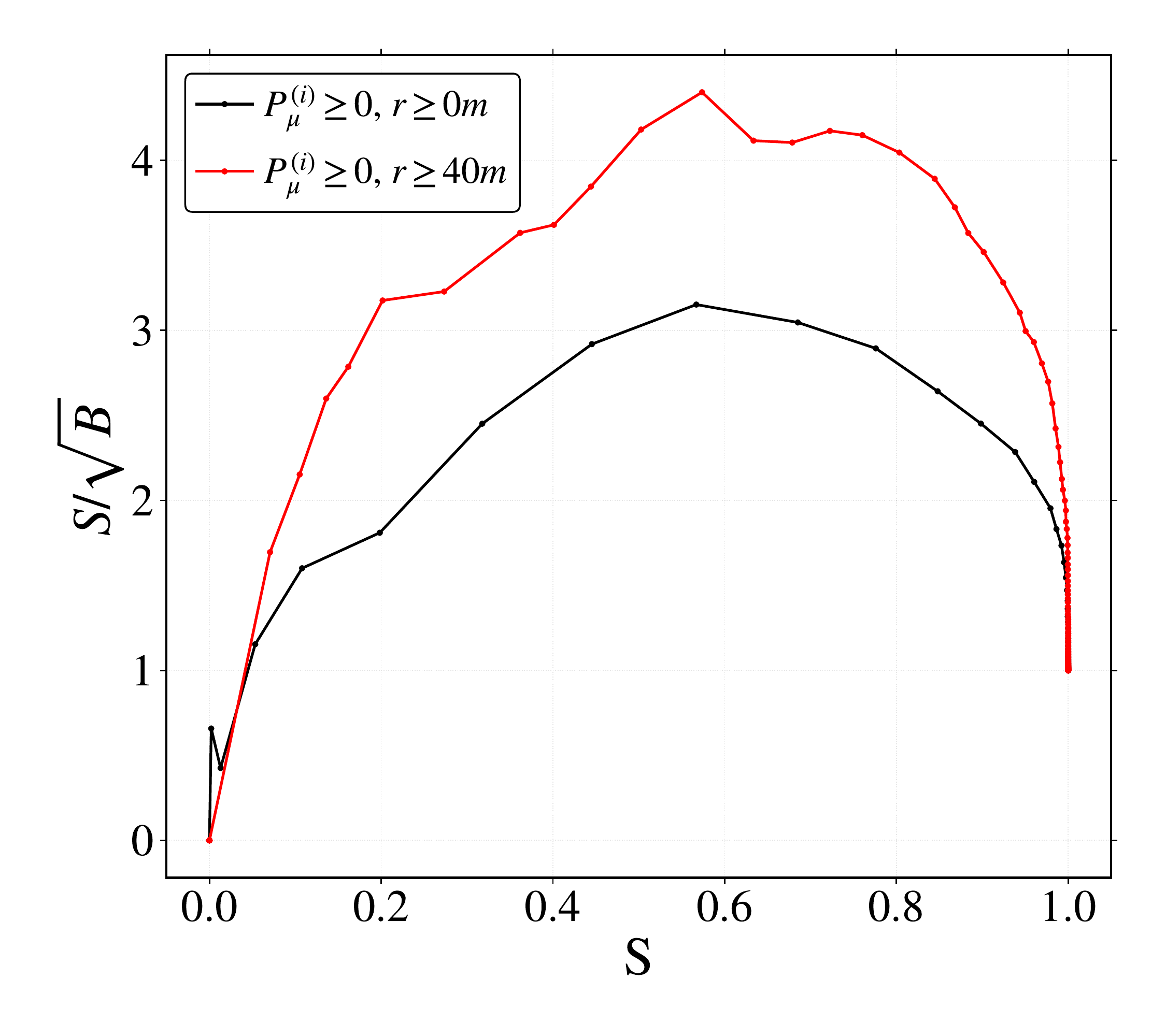}    
 \caption{Evaluation of the $S/\sqrt{B}$ achieved using selected stations as a function of the selection efficiency \textit{S} for gamma-rays.}
 \label{fig:sensitivity}
\end{figure}

In Figure \ref{fig:sensitivity} it is shown the $S/\sqrt{B}$ of the simulated instrument using this method. The motivation for sampling only stations far away from the shower core arises from shower physics considerations. On the one hand, muons can be produced in hadronically developing showers with high transverse momentum, and consequently falling far away from the shower core. On the other, the bulk electromagnetic shower component is much higher near the shower core. A cut on the probability $P_\mu^{(i)}$ was also tried without much difference on the results. A $S/\sqrt{B}$ of roughly 4 was found when fixing the selection efficiency for gammas to $S=0.6$. The obtained value is similar to the one quoted in other experiments such as LATTES\footnote{Simulation study in similar conditions.} \cite{LATTES} and HAWC \cite{HAWC_tank}, but using a significantly smaller WCD than HAWC. 

\section{Muon counting in proton-induced events} \label{sec:muon_counting}
In the previous section, a gamma/hadron separation was pursued. For that, it was enough identifying a certain quantity of the muons that reached stations. Nevertheless, muon counting capabilities when analysing hadronic induced EAS are of great interest. 

\begin{figure}[t]
 \centering
 \includegraphics[width=0.45\textwidth]{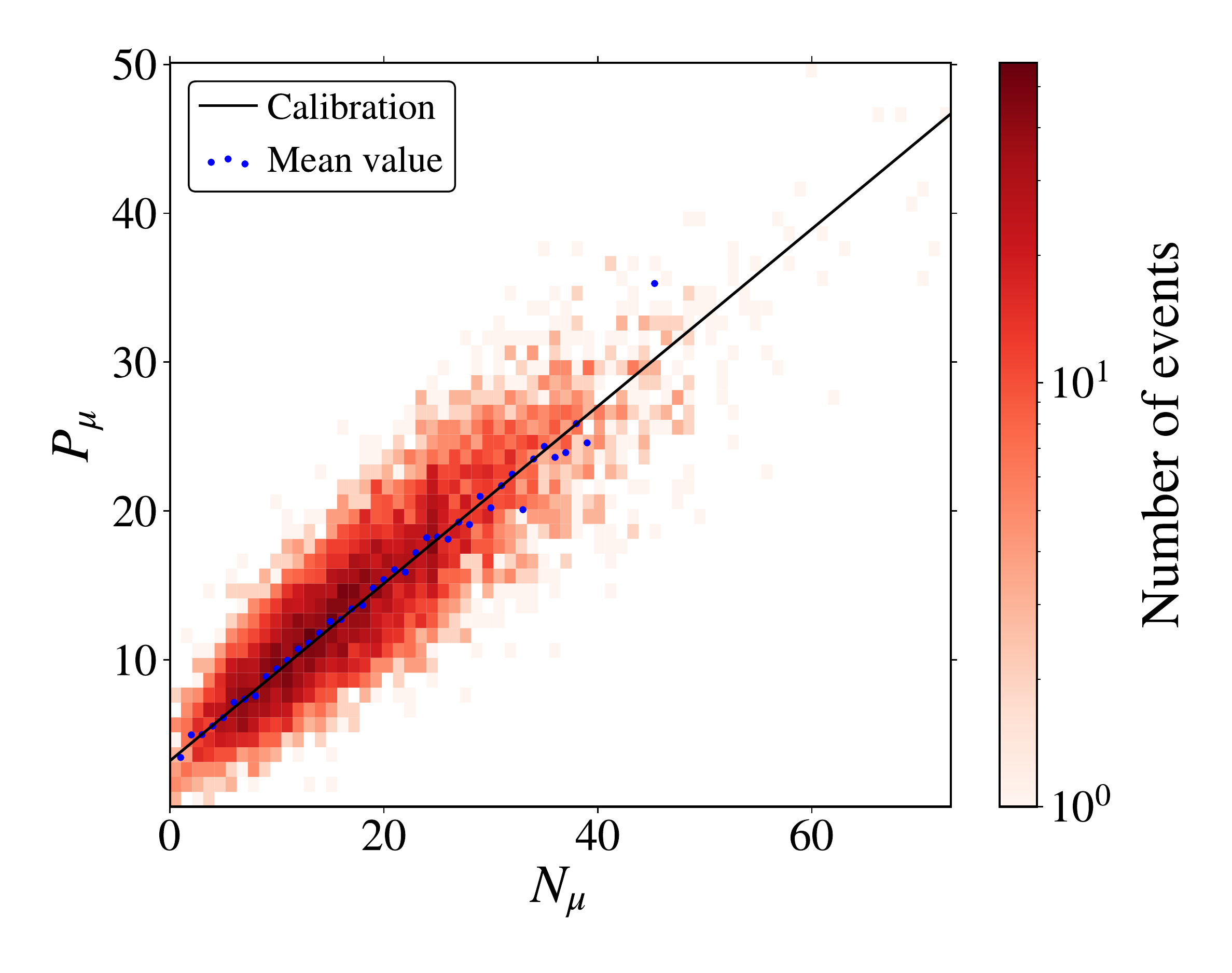}    
 \caption{Calibration of the relationship between $N_\mu = \sum_{i=0}^{N_S} N_{\mu}^{(i)} $ and $P_\mu = \sum_{i=0}^{N_S} P_{\mu}^{(i)}$ using proton induced air showers.}
 \label{fig:calibration}
\end{figure}

To prove that the discrimination achieved in Sections \ref{sec:muon_disc} and \ref{sec:gamma_hadron} comes from the fact that we are measuring muons and not any other shower feature, we decided to plot the correlation between $P_\mu$ and the number of muons crossing through the WCDs stations $N_\mu$ (this time we include events whose signal is lower than 300 p.e.) for each proton shower event. The result is shown in Figure \ref{fig:calibration} and it can be seen that there is a linear relationship between both variables. It should be noted that the ML method was not trained to count muons but instead to simply find them. However, for proton showers with energies $E_0 \sim 1\,$TeV the number of WCD stations with more than one muon is negligible, and as such, $P_\mu$ is in this conditions a good proxy for $N_\mu$.

It is then possible to calibrate it and estimate the number of muons by measuring the probability of having muons in all the stations. As such, the average probability $P_\mu$ assigned by the method to showers with the same number of muons $N_\mu$ is computed (see Figure \ref{fig:calibration}). Due to the scarce number of events with $N_\mu \geq 40$, an unique weighted mean value is considered taking into account all of them. We must take into account that, since we need to identify all the muons, no cuts on the distance to the shower core or on the probability $P_\mu$ are applied during this analysis.

Finally, in equation (\ref{eq:calibration}) it is described the calibration of the estimated number of muons $\hat{N}_{\mu}$ as a function of the amount of probability $P_\mu$ measured. The intercept of the calibration could be understood as the average probability from the stations without muons. In fact, this value $\sim 3$ is similar to the mean probability $P_{\mu}$ found in gamma-ray induced events of similar signal at the ground (see Figure \ref{fig:prob_stations}) where the presence of muons is very scarce.

\begin{equation} \label{eq:calibration}
    \hat{N}_{\mu} = 1.67 \cdot P_\mu - 3.22
\end{equation}

\begin{figure}[t]
 \centering
 \includegraphics[width=0.4\textwidth]{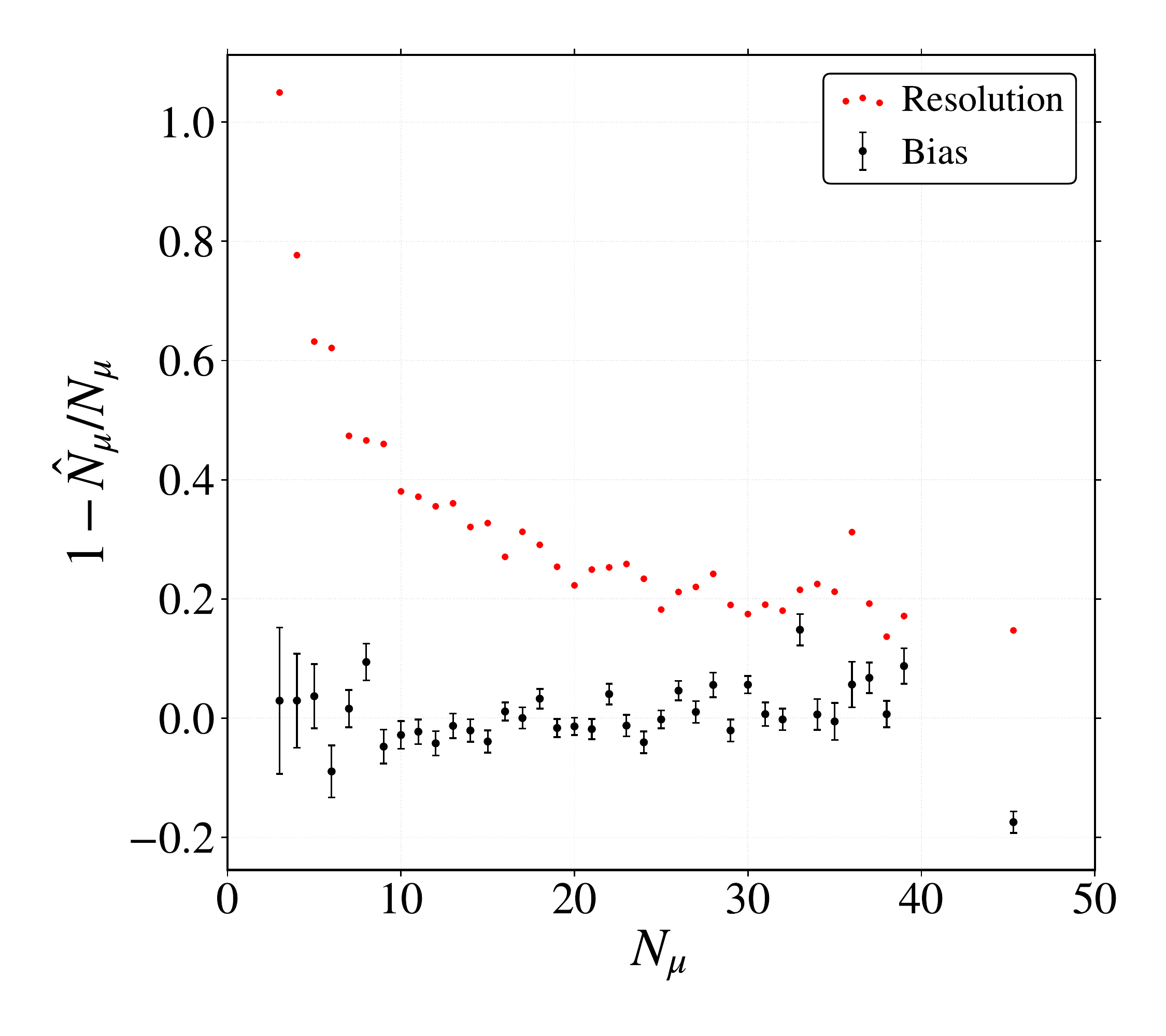}    
 \caption{Resolution and bias of the calibration between $N_{\mu}$ and $P_{\mu}$.}
 \label{fig:bias}
\end{figure}

In Figure \ref{fig:bias} it is shown both the resolution and the bias of the calibration proposed in equation (\ref{eq:calibration}). For events with $N_\mu > 10$ resolutions of about a $20 \%$ and a bias of just a few $\%$ were found. The resolution of the method for events with $N_{\mu} \geq 20$ can be adjusted to a function $1.06/\sqrt{N_{\mu}} + 0.02$. Therefore, the intrinsic resolution of the method is estimated to be just a $2 \%$. In view of these results, it can be seen that the samples with few muons are dominated by the electromagnetic noise, then larger fluctuations are found. As a consequence, as we increase the number of stations with muons, the electromagnetic contribution becomes a small portion of the overall probability $P_\mu$, which leads to more precise measurements. 

\section{Performance to inclined events and sparse arrays} \label{sec:discussion}
The proposed method has proven to be effective when identifying muons under the same circumstances used during the training stage: $E_0 \sim 4$ TeV and $\theta_0 \sim 10^{\circ}$. However, the possible degradation of the method performance in other scenarios must be discussed. An important factor is the inclination of the events and its impact in the results depending on the array configuration (dense or sparse).

\begin{figure}[t]
 \centering
 \includegraphics[width=0.45\textwidth]{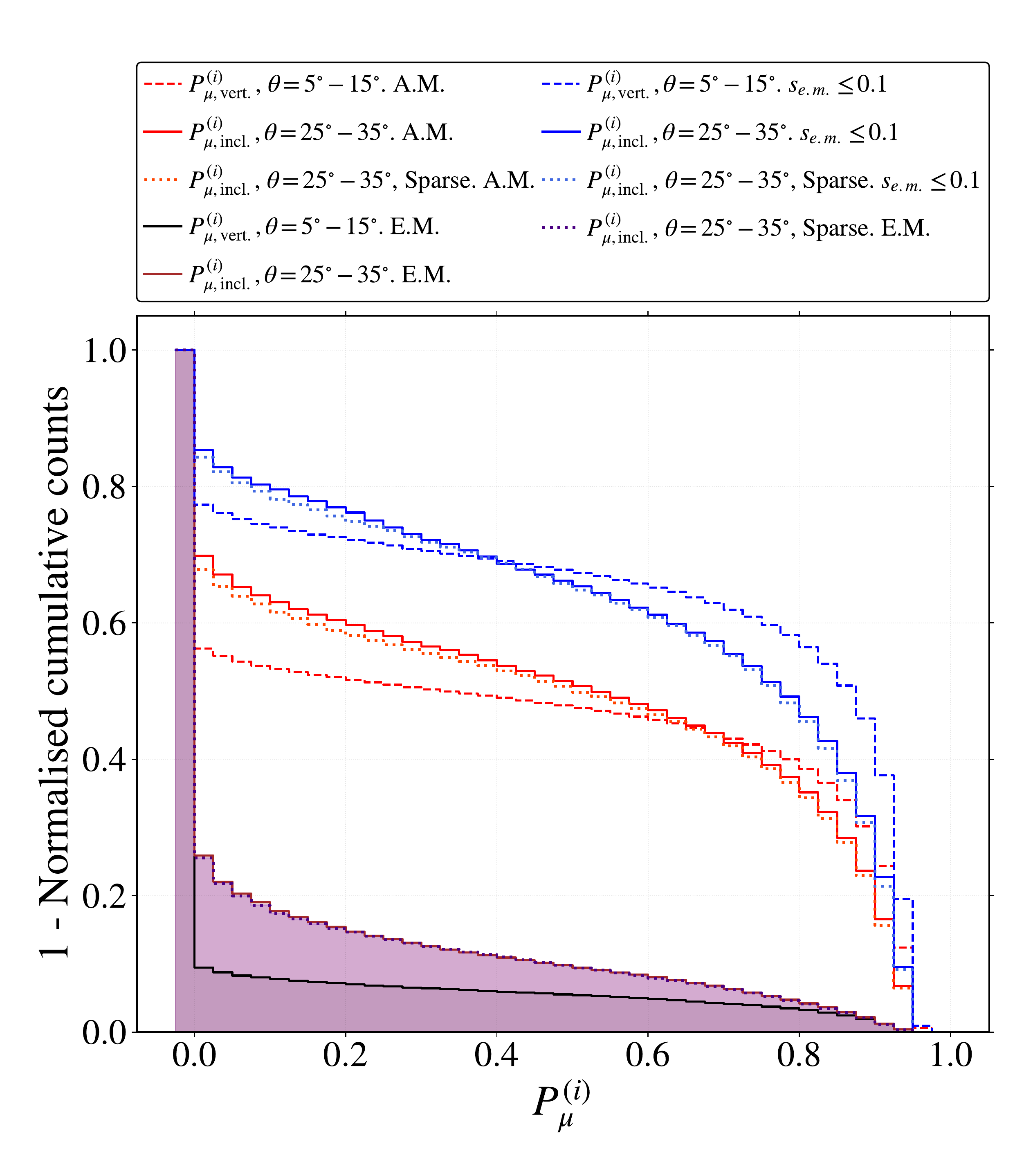}    
 \caption{Normalised inverse cumulative function of the probability $P_{\mu}^{(i)}$ of stations from proton-induced events with $\theta_0 \in \left[ 5^{\circ} ; 15^{\circ} \right]$ (dense array) and $\theta_0 \in \left[ 25^{\circ} ; 35^{\circ} \right]$ (dense and sparse array).}
 \label{fig:hist_sparse}
\end{figure}

To assess the impact of the shower zenith angle $\theta_0$ in the muon tagging, proton-induced events with $\theta_0 \in \left[ 25^{\circ} ; 35^{\circ} \right]$ were tested. Using the algorithm optimised for vertical showers it was found that roughly a $25 \%$ of All Muons had a probability $P_{\mu}^{(i)} \geq 0.5$. The result for stations without muons was similar to the one achieved using vertical showers. In view of this result, a CNN with the same configuration used for vertical showers was trained again using only inclined events (see Table \ref{tab:training_description} for more details). 

In Figure \ref{fig:hist_sparse} it is summarised all the tests done to vertical and inclined events in dense or sparse array. The normalised inverse cumulative function of the probability $P_{\mu}^{(i)}$ is shown for different combinations: inclined showers (using the algorithm that was trained for these showers $P_{\mu, \mathrm{incl.}}^{(i)}$) and vertical showers (using the algorithm previously trained with vertical showers $P_{\mu, \mathrm{vert.}}^{(i)}$).
The newly trained algorithm had a very similar performance for both arrays. 
Therefore, one can conclude that the performance of the muon tagging for these WCDs is not affected by the distance of the neighbour stations.

The proportion of each station class with probability $P_{\mu}^{(i)} \geq 0.5$ was: $\sim 70 \%$ of slightly contaminated muons; $\sim 55 \%$ of All Muons; and $\sim 10\%$ of the stations without muons. In comparison with the vertical events, a slight improvement was found when identifying All Muons whilst the performance with the stations without muons worsened.

The above results prove that it is possible to perform muon tagging using this WCD whether the shower event is vertical or inclined. Moreover, it also shows that the proposed WCD and method to tag muons can be applied whether the stations are close to each other or scattered in the field, making it a potential candidate to compact arrays, targeting the lower energies, and sparse arrays which try to reach the highest possible energies.

For the latter to be possible, it is our feeling that more work is necessary. Although we have not studied the higher energies (above $100\,$TeV), the electromagnetic contamination is expected to be higher, which might undermine the ability of tagging muons. As shown in Figure~\ref{fig:hist_contamination}, high electromagnetic contamination reduces the ability to tag muons. The electromagnetic component at high energies could still be dealt with by choosing WCD stations far away from the shower core position or removing it with statistical methods. Furthermore, the study of the higher energies would also require the use of a bigger detector array, which could be done by adding a sparse array surrounding the current compact array used in this paper. Therefore, such a study is out of the scope of the current work.

Finally, while the present work clearly demonstrates the possibility to tag/count muons with a relatively \emph{small} WCD through the exploration of the PMT asymmetry, its exact dimensions and number of PMTs should be obtained, balancing the physics performance (obtained through an end-to-end simulation) and its cost which shall depend on non-trivial factor such as site altitude, water availability, type of PMTs, among others.




\section{Conclusions} \label{sec:conclusions}

In this work, we have shown that a WCD with a water volume of $12\,{\rm m^2} \times 1.7\,{\rm m}$ and 4 PMTs can be used to efficiently tag muons. For that, a machine learning-based analysis, which processes the PMT acquired signals, has been developed. The analysis has been developed using WCD events solely from proton-induced showers with $E_0\sim 4\,$ TeV and zenith angles of $\theta_0 \sim 10^\circ$ for nearly vertical showers and $\theta_0 \sim 30^\circ$ for inclined showers. 

In this paper, we show that the identification of the muon in the station, given as a probability, $P_\mu^{(i)} \in \left[ 0;1 \right]$, depends on the amount of electromagnetic contamination in the station, with nearly no dependence on the configuration of the WCD array (dense vs sparse). The proposed method was effective for both vertical and inclined induced events, where roughly the $70 \%$ of stations with slightly contaminated muons and the $50 \%$ of all stations with muons had $P_\mu^{(i)} > 0.5$. For inclined events, just 10\% of the stations without muons had $P_\mu^{(i)} > 0.5$, while for vertical events the same happens for $P_\mu^{(i)} > 0$.



Using a simple gamma/hadron discrimination observable created from  $P_\mu^{(i)}$ it has been shown that it is possible to distinguish between gamma and proton-induced showers with a $S/\sqrt{B} \sim 4$ for shower energies of $E_0 \sim 1\,$TeV and $\theta_0 \sim 10^\circ$. Moreover, it has been shown that a calibration between $P_\mu$ and $N_\mu$ can be built, leading to a negligible bias for $N_\mu > 10$ and resolutions of $\sim 20\%$.

\begin{acknowledgements}
We would like to thank to A. Bueno for all the support and useful discussions during the development of this work.
The authors thank also for the financial support by OE - Portugal, FCT, I. P., under project PTDC/FIS-PAR/29158/2017.
R.~C.\ is grateful for the financial support by OE - Portugal, FCT, I. P., under DL57/2016/
cP1330/cT0002. 
B.S.G. is grateful for the financial support by grant LIP/BI - 14/2020, under project IC\&DT, POCI-01-0145-FEDER-029158. 
\end{acknowledgements}

\bibliography{references.bib}   

\end{document}